\documentclass[12pt]{iopart}

\begin{document}
\newcommand{\eq}{\begin{equation}}
\newcommand{\eqx}{\end{equation}}
\newcommand{\be}{\begin{eqnarray}}
\newcommand{\ee}{\end{eqnarray}}
\newcommand{\f}[2]{\frac{#1}{#2}}
\newcommand{\lm}{\lambda}

\newcommand{\al}{\alpha}
\newcommand{\om}{\omega}
\newcommand{\sg}{\sigma}
\newcommand{\eps}{\varepsilon}
\newcommand{\dl}{\delta}
\renewcommand{\AA}{{\cal A}}
\newcommand{\TT}{{\cal T}}
\newcommand{\AAd}{{\cal A}^\dagger}
\newcommand{\gm}{\gamma}
\newcommand{\Gm}{\Gamma}
\newcommand{\OM}{\Omega}
\renewcommand{\SS}{{\cal S}}

\newcommand{\arr}[4]{%
\left(\begin{tabular}{c|c}
$#1$ & $#2$ \\
\hline $#3$ & $#4$
\end{tabular}\right)}

\renewcommand{\det}{\mbox{\rm det}}
\title[Random walkers versus random crowds]{Random walkers versus random crowds: diffusion of large matrices}
\author{Ewa Gudowska-Nowak$^1$, Romuald A. Janik$^1$, Jerzy Jurkiewicz$^1$, Maciej A. Nowak$^1$, Waldemar Wieczorek$^2$}
\address{$^1$ M. Smoluchowski Institute of Physics and M. Kac Complex Systems Research Center, Jagiellonian University,
 Reymonta 4, 30-059 Krak\'ow,
Poland}
\address{$^2$ Fachbereich Physik, Universit\"at Duisburg-Essen, 47048 Duisburg, Germany}

\ead{gudowska@th.if.uj.edu.pl}
\begin{abstract}
We briefly review the random matrix theory for large $N$ by $N$
matrices viewed as free random variables   in a context of
stochastic diffusion. We establish a surprising link between the
spectral properties of matrix-valued multiplicative diffusion
processes for  hermitian  and unitary  ensembles.
\end{abstract}

\maketitle
\section{Introduction}
Random matrix theory (RMT) represents a powerful tool in several
statistical problems where the relevant degrees of freedom can be
encoded as elements of certain ensembles of large matrices.
 Applications cover practically all branches of theoretical physics (spectral
properties of excited nuclear, atomic and molecular systems,
simplicial gravity, theory of mesoscopic conductance or Euclidean
Quantum Chromodynamics) and are more and more frequent in recent
interdisciplinary research. Indeed, contemporary complex systems
are characterized by massive storage of data, usually cast in the
form of matrices. Typical examples are: huge covariance matrices
for financial and/or economic indicators, incidence matrices for
information/retrieval search engines, adjacency matrices for
networks, stochastic matrices describing frequencies of alleles in
various gene populations, frequency response matrices in wireless
telecommunication systems, to mention only few. Recently, concepts
of free random calculus~\cite{VOICULESCU,SPEICHER} allowed  to
build in parallel to Classical Probability Calculus(hereafter CPC)
a new framework -- {\em matrix-valued probability theory},
hereafter named Free Probability Calculus (FPC), and allowed to
recast several mathematical features of large size random matrices
in an intuitive and user-friendly way.

Most of the above-mentioned applications of random matrices of
free random variables  correspond to the "static case", when we
are only interested in spectral properties of certain ensembles,
and not in the dynamical properties of evolution of such ensembles
as a function of some external parameter, representing either real
time or  the length
 of a disordered wire, temperature  or the numerical value
of the coupling constant. In all these realms matrix ensembles are
becoming evolving dynamical systems.  This is the problem we are
addressing  in this note: using the parallel between standard
probability calculus for one-dimensional random variables and
noncommutative probability calculus of large matrices (FPC) we
consider the simplest matrix-valued stochastic evolutions -
Brownian walks. Note, that in our case the stochastic increment
is represented by a whole matrix, so we deal rather with Brownian
crowds than with single walker.

The formulation of this problem was addressed in our previous
papers~\cite{DIFF1,DIFF2,DIFF3}, where we have also given
references to related literature on this subject. Here, after very
brief familiarization with such  concepts of free variables as
multiplicative S transformation (section~2), we review two simple
cases of stochastic multiplicative evolutions considered recently
in the literature~\cite{DIFF1,DIFF3} (unitary multiplicative
evolution and hermitian  multiplicative evolution)  and we
establish a new link between their spectral properties.

\section{S transformation}

One dimensional diffusion process for geometric random walk (in
the absence of drift) is given by the stochastic differential
equation \be \frac{dy}{y}= \sigma dx_{\tau} \ee where $y$ is
random and $dx_{\tau}$ represents Wiener measure. By averaging
$y(\tau)$ we recover well known solution for the probability
density, so-called log-normal law \be p(y,\tau|
y_0,0)=\frac{1}{y\sqrt{2\pi \sigma^2 \tau}} \exp \left[ -
\frac{(\log(y/y_0)+\frac{1}{2} \sigma^2\tau)^2}{2\sigma^2
 \tau} \right]
\ee

The corresponding analogue in the space of matrices may be
 written as the
product of random matrix-valued variables \be Y_{\tau}=\prod_i^K
\left( 1+\sqrt{\frac{\tau}{K}}X_i\right) \label{def.Y} \ee where
$X_i$ represent large (infinite) hermitian Gaussian free matrices
belonging to GUE with the unit variance and infinitesimal time
step is equal to $\tau/K$.

To solve this  multiplicative diffusion process we notice, that
since matrices are non-commuting objects, a standard
exponentiation trick relating algebraic to geometric Brownian walk
fails. This is why we have to use the concepts of free random
variables. We proceed in the following steps.

First, we  consider an   ensemble built of large $N \times N$
matrices $M$,
 for which
the pertinent resolvent (Green's function) is generically given by
\be G(z) = \int dM  e^{-N\,V(M)}\frac 1N\,\, {\rm Tr}\left( \frac
1{z-M}\right)\,, \label{0} \ee In the case of Gaussian ensemble,
potential $V(M)$ equals $\frac{1}{2} {\rm Tr} M^2$. For large $z$,
the Green's function is the generating function for spectral
moments~(we do not consider here so-called L\'{e}vy ensembles, for
which moments do not exist). In the Gaussian case
$G(z)=(z-\sqrt{z^2-4})/2$, so the imaginary part yields Wigner's
semicircle spectral distribution and expansion in $1/z$ generates
all the spectral moments of this distribution.

Second, note that the product of hermitian matrices is, in
general, not a hermitian matrix anymore, which means that the
eigenvalues start to diffuse over the whole complex plane, and do
not stay on the real axis. This problem was recently solved by us
using diagrammatic tools of random matrix theory~\cite{DIFF1}.
Note however, that in the case when we are only interested in {\it
moments} of the resulting non-hermitian distribution, we may still
use the concepts of the generating functions, i.e. resolvents, and
considerably reduce the technical complexity of the whole problem.
This is due to the fact that even for non-hermitian ensembles,
whose eigenvalues lie in bounded domains of the complex plane, the
Green's function (\ref{0}) is analytic around $z=\infty$ and thus
serves as a generating function for the moments.

In order to carry out this program in practice, we use  here a
remarkable tool of FPC known as multiplicative S transformation,
allowing us to find a generating function for all moments of the
product, provided the spectral properties of the individual
ensembles are known. The operational procedure is quite simple:
Let us assume that we know Green's $G_i(z)$ functions for each of
the hermitian ensembles $M_i$. We construct an auxiliary
analytical function $\chi_i$ for each of them, defined by the
condition \be \frac{1}{\chi_i} G_i\left( \frac{1}{\chi_i} \right)
-1=z \ee and then we redefine the result, getting
$S_i(z)\equiv\frac{1+z}{z} \chi_i(z)$. The resulting S-transform
is {\em multiplicative}, i.e. \be S_{1 \times 2}(z)= S_1(z) \cdot
S_2(z) \ee and represents {\it the multiplication law} for the
spectral cumulants.

Applying the concept to individual matrix-valued entry
$1+\sqrt{\tau/K}X$, we obviously recover Wigner's semicircle
shifted by 1 and with rescaled variance $\beta=\tau/K$. A short
calculation~\cite{DIFF2} yields the desired  S-transformation for
the product \be S_Y(z)=\lim_{K \rightarrow \infty} S_K =\lim_{K
\rightarrow \infty}
 \prod_K S_i
= \lim_{K \rightarrow \infty}  \left(\frac{-1 + \sqrt{1+4\tau
    z/K}}{2\tau z/K}\right)^K
\ee The surprisingly simple answer reads
 \be S_Y(z)=\exp(-\tau z)
\label{masterS} \ee Note that despite the fact that we have
considered initially only Gaussian  ensembles (with unit
variance), the above result is valid for any hermitian random
matrix ensemble provided its second moment is finite and
normalized to 1. In such a way it represents a new central limit
theorem for the product of infinitely many hermitian entries.

Let us now consider a quite different multiplicative evolution,
solved recently by two of the present authors~\cite{DIFF3}, namely
the spectrum of a unitary operator, represented by an infinite
product of unitary, free (independent) increments. Now the
evolution operator is represented as \be Y_{\tau}=\prod_i^K \exp
\, \left( i\sqrt{\frac{\tau}{K}} H_i \right) \ee where one assumes
only that the second spectral moments of hermitian $H_i$ are
finite. Note that in this case the spectrum always stays on the
unit circle, since the product of unitary matrices is unitary. In
this case, the S-transformation is again surprisingly simple and
the result reads~\cite{DIFF3} \be S(z)=\exp(\tau (z+1/2))\ee where
we have set the arbitrary variance of the initial distribution to
one, to simplify the comparison between the unitary and hermitian
cases considered here.

\section{Dynamical evolution of both ensembles}

 To distinguish between both cases of unitary and hermitian S
 transformations, we denote them by $U$ and $H$ subscripts,
 respectively. Since $S_H(z)=\exp (-\tau z)$ and $S_{U}(z)=\exp \tau(
(z+1/2)$, we notice that \be S_H(z)S_U(z)=e^{\tau/2} \label{dual}
\ee i.e. spectral properties of moments are very closely related.
To see the relation to moments  in more detail, we  recall that
the Green's function $G(z)$ for the unitary diffusive matrix
ensembles~\cite{DIFF3} generates moments \be \frac{1}{z} G
(\frac{1}{z}) -1=\sum_{n=1}^{\infty} m_n z^n \ee where \be
m_n(\tau) =< {\rm Tr } Y_{\tau}^n>=\frac{1}{2\pi}\int \rho(\theta,
\tau)e^{in \theta} d \theta \label{moments} \ee Since  the
multiplicative Voiculescu S transformation is related to the
functional inverse of the generating function: \be
\frac{1}{\chi(z)} G[\frac{1}{\chi(z)}]-1=z \ee where
$S(z)=\frac{1+z}{z} \chi(z)$, one can apply standard residua trick
for inverting the series~\cite{MF}, getting  that \be
m_n=\frac{1}{n} \oint \frac{dz}{2\pi i} [\chi(z)]^{-n} \ee In the
case of $S_U(z)$ the integral representation for moments reads:
\be m_n=\frac{1}{n} \oint \frac{dz}{2\pi i}(1+1/z)^n
\exp(-n\tau(z+1/2))= \frac{1}{n}e^{-n\tau/2} L_{n-1}^{(1)} (\tau
n) \ee where we used the integral representation of the Laguerre
polynomials. This result resembles the structure observed in the
Gross-Witten type models~\cite{GW}. A closer look shows, that the
similarity to Gross-Witten models or string models is actually
{\it exact}, provided that the corresponding coupling constant in
this model (or the surface of the string in string models) is
identified with the evolution parameter $\tau$ (in units where the
second spectral moment is one). As far as we now, this simple
observation relating the generic models to {\it diffusive}
properties of large matrices was never explicitly stated in the
literature.

\section{Phase transitions}

It is important to notice that both considered models have phase
transitions. In the case of unitary diffusions, the phase
transition is visible e.g.  from the large $n$ behavior of
moments. In this case,
 asymptotics of Laguerre polynomials known as Plancherel-Rotach
limit confirms the critical
 value $\tau_c=4$, above which  moments decay exponentially and  for times smaller than critical,
moments oscillate with amplitude $n^{-3/2}$. These results were
explicitly derived in the context of two-dimensional  QCD by
Gross and Matytsin~\cite{GM}, provided we  identify string surface
$A$ with $\tau$. Resulting phase transition is third order
(Douglas-Kazakov type), and generic for all types of Gross-Witten
type models. It corresponds to a geometrical phase transition,
when {\it diffusing} spectra, originally concentrated at single
point or the finite arc of the unit circle, spread in both
orientations  of the circle to meet and to close the gap
previously free of eigenvalues. Our duality relation (\ref{dual})
suggests the presence of phase transition at $\tau_c=4$ also for
the case of the the product of hermitian ensembles.  This was the
case observed analytically and confirmed numerically in
\cite{DIFF1}.  The phase transition there was of "topological
nature" again - it corresponded to the appearance of the hole on
the plane of diffusing {\it complex} eigenvalues.

\section{Burgers equations}

Last but not least, we address the issue of dynamical equations
describing above-mentioned phenomena. To understand the structure
of the {\em a-priori} unknown  equations, we consider first the
simple additive evolution of Wigner semicircle. This is the case
when the length of support averaged spectral distribution of
eigenvalues has to grow as $\sqrt{\tau}$, in analogy to
Einstein-Smoluchowski formula for "classical diffusion".  This
process was already considered in 1962 by Dyson~\cite{DYSON}, who
was considering stochastic differential Langevin equations for
temporal behavior of the eigenvalues \be
d\lambda_i(\tau)=\frac{1}{\sqrt{N}}\,dx^{(i)}_{\tau} +
\frac{1}{N}\,\sum_{i \neq j}
\frac{1}{\lambda_i(\tau)-\lambda_j(\tau)} dt \ee where the first
term represents Wiener process and the second, originating from
the measure (van der Monde determinant) represents deterministic,
strong repulsive force between any pair of eigenvalues.  For large
$N$ individual Langevin-like equations  may be represented by the
equation for average spectral distributions $\rho(\lambda,\tau)$,
i.e.  analogs of Fick/Fokker-Planck equations known in classical
case. They read~\cite{SPEICHERBIANE}

\be \partial_{\tau} \rho(\lambda,\tau) +\partial_{\lambda} \left[
\rho(\lambda,\tau) {\rm PV} \int
\frac{\rho(\lambda^{'},\tau)}{\lambda-\lambda^{'}}d \lambda
\right]=0 \ee therefore are {\em non-linear}. An alternative way
to recast these equations is to use the Green's function. Since
imaginary part of the Green's functions is  related to spectral
density, matrix-valued  "diffusion equations" take the form \be
\partial_{\tau}G(z,\tau) + G(z,\tau)\partial_zG(z,\tau)=0
\label{Bur.triv} \ee known also as complex  Burgers equation.

We can return now to the case of the hermitian multiplicative
diffusion. Knowing $S_H(z)=\exp (-\tau z)$ we  read out the
Green's function for the ensemble $Y$ \be
G_Y(z,\tau)=\frac{1+f(z,\tau)}{z} \ee where $f$ is the solution of
the transcendental equation \be zf=(1+f)e^{\tau f} \label{imp} \ee
Note, that by definition, $G_Y(z,\tau)$ is a holomorphic function
(defined for large~$z$) generating {\it spectral moments} of
$Y_{\tau}$, i.e. \be G_Y(z,\tau)=\sum_{n=0}^{\infty} {\rm tr\,\,}
Y_{\tau}^n \frac{1}{z^{n+1}} \label{holGreen} \ee

Let us remind~\cite{DIFF2} that $S_Y$ fulfills a simple equation
\be z\partial_z S-\tau\partial_{\tau} S=0 \ee By differentiating
(\ref{imp}) with respect to $\tau$ and $z$ we get the following
evolution equation \be
\partial_{\tau} f + zf\partial_z f=0
\label{logBur} \ee or equivalently \be
\partial_{\tau} f +f \partial_{\ln z}f=0
\ee supplemented by the boundary condition
$f(\tau=0,z)=(z-1)^{-1}$. This is the Burgers equation where
spatial evolution is governed by $\log z$. Note however, that this
time the semicircle solution of the additive Burgers equation
(\ref{Bur.triv}),
 with $z$ replaced
by $\ln z$, {\it does not} fulfill (\ref{logBur}). It should not
be puzzling for the reader that similar Burgers-like equations
describing the evolution of the spectra one encounters  also in
the case of unitary evolutions~\cite{DIFF3,GM}.

\section{Summary}

In this paper we discussed some problems of Brownian-like walks in
the space of the spectra of large matrices with the help of
powerful tools of free random variables calculus. New results
presented in this note include: (i) a rather unexpected link
between the moments of unitary evolutions and hermitian
evolutions; (ii) an exact correspondence  between diffusive
character of  evolving large unitary matrices and the whole class
of generic models of Gross-Witten type known in two-dimensional
Quantum Chromodynamics. We hope that the language of matrix-valued
evolution presented here does not only form an exciting and rich
mathematical structure, but may also help to understand  basic
concepts of  "new statistical physics" of large and non-commuting
objects.

\subsubsection*{Acknowledgments}

This work was supported in part by the Polish State Committee for
Scientific Research (KBN)  grants 2P03B08225 (2003-2006), 1P03B
02427 (2004-2007) and 1P03B04029 (2005-2008) and the Marie Curie
Actions Transfer of Knowledge project COCOS (contract
MTKD-CT-2004-517186). Waldemar Wieczorek acknowledges a financial
support within the DFG programme SFB/TR12. The authors are
grateful to Roland Speicher and Piotr \'{S}niady for valuable
remarks on Free Random Variables calculus, and to Poul Olesen for
the correspondence on the link with 2D YM.

\newpage

 {\bf References}\\

\end{document}